%% file: main.tex
\documentclass{article}
\usepackage{spconf,amsmath,graphicx}
\usepackage{url}
\usepackage{subfigure}
\usepackage{amssymb}
\usepackage{multirow}
\usepackage{threeparttable}
\usepackage{xcolor}
\usepackage{float}
\usepackage{graphicx}

\title{On Efficient Neural Network Architectures for Image Compression}

\name{Yichi Zhang, Zhihao Duan, Fengqing Zhu}
\address{Elmore Family School of Electrical and Computer Engineering,\\Purdue University, West Lafayette, Indiana, U.S.A.}

\begin{document}
%
\maketitle
\begin{abstract}
Recent advances in learning-based image compression typically come at the cost of high complexity. Designing computationally efficient architectures remains an open challenge. In this paper, we empirically investigate the impact of different network designs in terms of rate-distortion performance and computational complexity. Our experiments involve testing various transforms, including convolutional neural networks and transformers, as well as various context models, including hierarchical, channel-wise, and space-channel context models. Based on the results, we present a series of efficient models, the final model of which has comparable performance to recent best-performing methods but with significantly lower complexity. Extensive experiments provide insights into the design of architectures for learned image compression and potential direction for future research. The code is available at \url{https://gitlab.com/viper-purdue/efficient-compression}.
\end{abstract}


\input{introduction.tex}

\input{related_work.tex}
\input{transform_study}

\input{context_study.tex}

\input{experimental_setup.tex}
\input{complexity.tex}

\input{conclusion.tex}

\bibliographystyle{IEEEtran}
{\footnotesize
\bibliography{Efficient.bib}
}
\end{document}

%% file: introduction.tex
\section{Introduction}
\label{sec:intro}

With the widespread use of high-resolution cameras and the growth of image-centric social media and digital galleries, images have become a dominant form of media in daily life. The substantial size of images, particularly those with high resolution, places considerable demands on transmission bandwidth and storage capacity. In the past decades, lossy image compression has become a particularly vital technique for high-efficiency visual communication. Many rule-based compression algorithms (e.g., JPEG~\cite{wallace1992jpeg}, JPEG2000~\cite{christopoulos2000jpeg2000}, HEVC/H.265~\cite{sullivan2012overview} intra, and VVC/H.266~\cite{bross2021overview} intra) have been proposed to improve the compression efficiency. 

Recently, learning-based image compression (LIC) methods have been highly regarded due to their simple framework and impressive performance. Some ~\cite{he2022elic,liu2023learned,zhang2024theoretical} have even surpassed the best-performing rule-based compression method, VVC/H.266~\cite{bross2021overview} intra, in terms of rate-distortion (RD) performance. However, one major concern with LIC methods is their complexity. While achieving high RD performance remains a priority, there is now an increasing interest in developing practical LICs that balance the rate, distortion, and complexity. This balance is commonly referred to as the Rate-Distortion-Complexity (RDC) trade-off~\cite{gao2023exploring,minnen2023advancing,guo2023cbanet}.

\begin{figure}[t] 
    \centering 
    \includegraphics[width=0.92\linewidth]{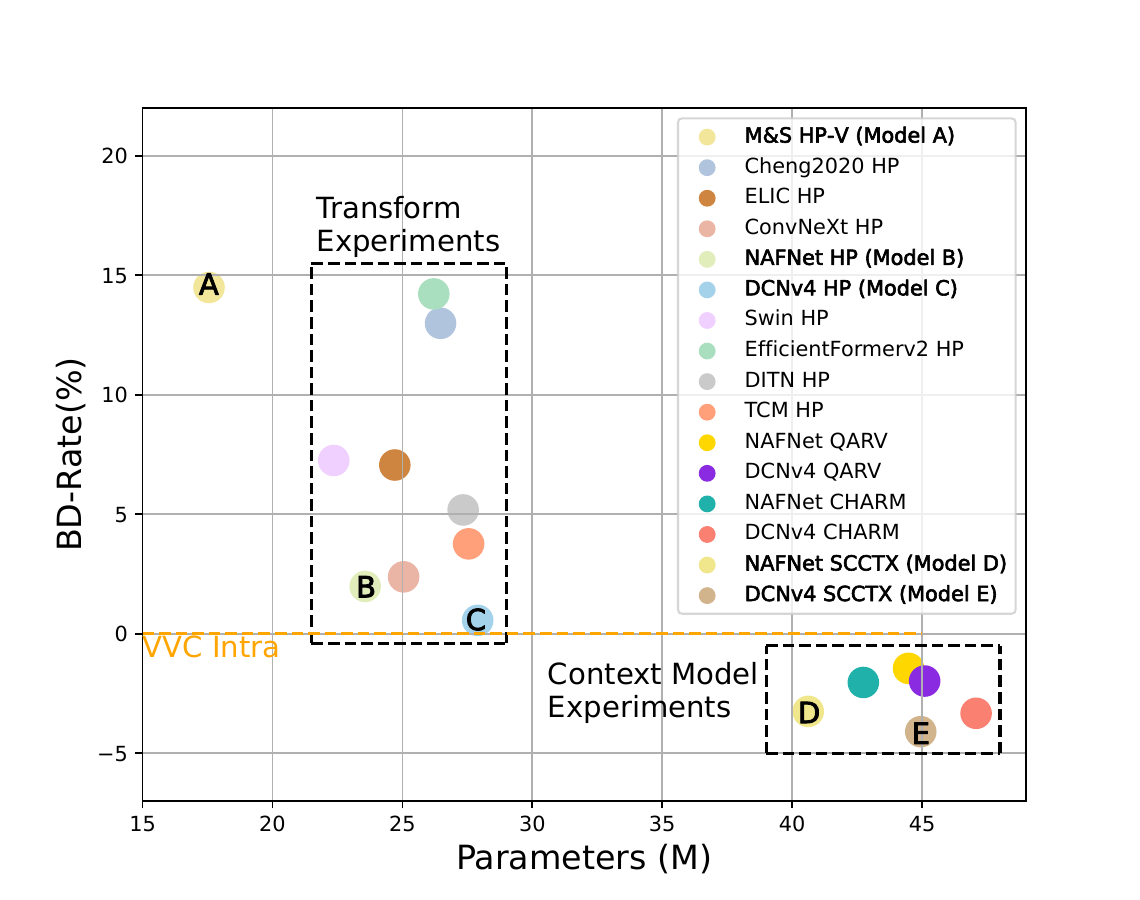}
    \vspace{-0.4cm}
    \caption{{BD-Rate (smaller is better) vs model parameters on Kodak.} Models A, B, C, D, and E represent the best RDC trade-off.}
    \label{fig:rdc}
\end{figure}

In this paper, we explore the RDC trade-off by conducting a thorough empirical study. We break down the LIC framework into transforms and context models and use model parameters as a measure of complexity along with other metrics (Section~\ref{sec:com}). 
Transforms map images to the latent space for decorrelation and energy compression, while context models improve coding efficiency by converting the marginal probability model of coding symbols into a joint model by incorporating additional latent variables as priors. Therefore, the key to designing efficient LICs lies in the synergistic combination of effective transforms and context models.

Our study begins with a variable rate baseline model inspired by the M\&S Hyperprior~\cite{minnen2018joint} and then explores various network structures for the transform. We then integrate advanced low-complexity context models and conduct extensive experiments on different combinations of transforms and context models. As a result, we identify a series of models that achieve the best RDC trade-offs, offering high performance with low complexity (\textbf{Contribution 1}). The performances of these models are illustrated in Fig~\ref{fig:rdc}. Additionally, we analyze these models in detail to gain insights into the contributing factors to their performance improvement (\textbf{Contribution 2}). Notably, Our final model, DCNv4 SCCTX, outperforms most existing methods in terms of RDC (\textbf{Contribution 3}). Each model in this study supports variable rates and represents an RD trade-off under specific complexity.

%% file: related_work.tex
\section{Related Work}
\label{sec:re}

\subsection{Transform}
Transforms serve as the core technology for contemporary compression methods. The JPEG standard~\cite{wallace1992jpeg} utilizes the discrete cosine transform (DCT)~\cite{ahmed1974discrete}, whereas JPEG2020~\cite{christopoulos2000jpeg2000} employs the discrete wavelet transform (DWT)~\cite{shensa1992discrete}. The advanced ITU-T H.26x series~\cite{sullivan2012overview,bross2021overview} primarily rely on the type-2 DCT (DCT-2)~\cite{karhunen1947under} which closely approximates the optimal data-driven Karhunen-Lo{\`e}ve transform (KLT) under the first-order Markov conditions~\cite{zhao2021transform}. In addition to these traditional methods, the development of deep learning has also led to the proliferation of LIC methods. Ball{\'e}~{\it et al.}~\cite{balle2016end} integrates stacked convolutions with Generalized Divisive Normalization (GDN) and inverse GDN (IGDN), for the transform function. Subsequent advancements introduce attention modules~\cite{cheng2020learned} and context clustering~\cite{zhang2024way} for image de-correlation. The rise of self-attention-based Transformers also prompts their exploration as transforms. Zhu~{\it et al.}~\cite{zhu2022transformer} employs the Swin Transformer while Zou~{\it et al.}~\cite{zou2022devil} utilizes a symmetrical Transformer. Recently, Liu~{\it et al.}~\cite{liu2023learned} combines the designs of convolutions and Swin Transformer resulting in improved performance.

\subsection{Context Model}
Context model plays a crucial role in improving entropy coding efficiency. In the ITU-T H.26x series~\cite{sullivan2012overview,bross2021overview}, techniques such as context-adaptive variable-length coding (CAVLC)~\cite{moon2005efficient} and context-adaptive binary arithmetic coding (CABAC)~\cite{sze2012high} leverage statistical correlations among coding symbols to reduce redundancy. Similarly, in LIC, context models are designed to encode symbols at the lowest possible bit rate. The spatial autoregressive context model~\cite{minnen2018joint} and its variants~\cite{qianentroformer} encode symbols sequentially. However, their serial processing nature leads to long coding times, making them less practical. To address this issue, parallel context modeling methods have been developed for better applicability in real-world scenarios. For example, the checkerboard model~\cite{he2021checkerboard} enables independent encoding of anchor contents, with non-anchor contents benefiting from their dependency on anchor content priors. Subsequently, the dual spatial prior model~\cite{guo2023evc} is introduced to further enhance this approach. Beyond spatial aspects, numerous works have been devoted to exploring the correlation across channels. The channel-wise autoregressive model (CHARM) is proposed to divide channels into slices and condition them on previously coded channels~\cite{minnen2020channel}. Observing the uneven distribution of information content among channels, He~{\it et al.}~\cite{he2022elic} introduces a space-channel context model (SCCTX), which integrates an uneven channel grouping strategy with the checkerboard model for improved performance. More recently, Duan~{\it et al.}~\cite{duan2023qarv} presents the Quantization-Aware ResNet VAE (QARV), a hierarchical model that can generalize autoregressive models, including the spatial autoregressive model~\cite{minnen2018joint} (for a constant image resolution), the CHARM~\cite{minnen2020channel}, and the SCCTX~\cite{he2022elic}.

\subsection{Rate-Distortion-Complexity Trade-off}
The Rate-Distortion-Complexity (RDC) trade-off has been a critical focus in the field of data compression. Hu~{\it et al.}~\cite{hu2006joint} introduces a framework within the H.264 standard that optimizes the balance between coding efficiency and computational complexity for motion search in embedded systems. Foo~{\it et al.}~\cite{foo2008analytical} conducts a comprehensive RDC analysis of wavelet video coding, incorporating a detailed model of several aspects found in operational coders, considering elements such as embedded quantization, quadtree structures for block significance mapping, and context-adaptive entropy coding for subband blocks. In recent years, the RDC trade-off in LIC also shines. Gao~{\it et al.}~\cite{gao2023exploring} systematically investigates the RDC trade-off, quantifying decoding complexity to fine-tune the balance. Minnen~{\it et al.}~\cite{minnen2023advancing} conducts a rate-distortion-computation frontier study that resulted in a family of model architectures, achieving an empirical balance between computational requirements and rate-distortion performance. 
\textbf{Remarks.} 
Our study is inspired by prior research~\cite{minnen2023advancing}, but different in several ways: 1) our models accommodate variable rates covering a wide range of bit-rates (0.16 bpp-2.8 bpp), 2) our models do not confine the transforms to the compression-related methods, as discussed in Section~\ref{sec:Transform}, 3) our models expand the evaluation of context models by introducing a greater variety of options, as elaborated in Section~\ref{sec:Context}.

%% file: transform_study.tex
\section{Transform architecture Study}
\label{sec:Transform}
\begin{figure}[htbp] 
    \centering 
    \includegraphics[width=0.85\linewidth]{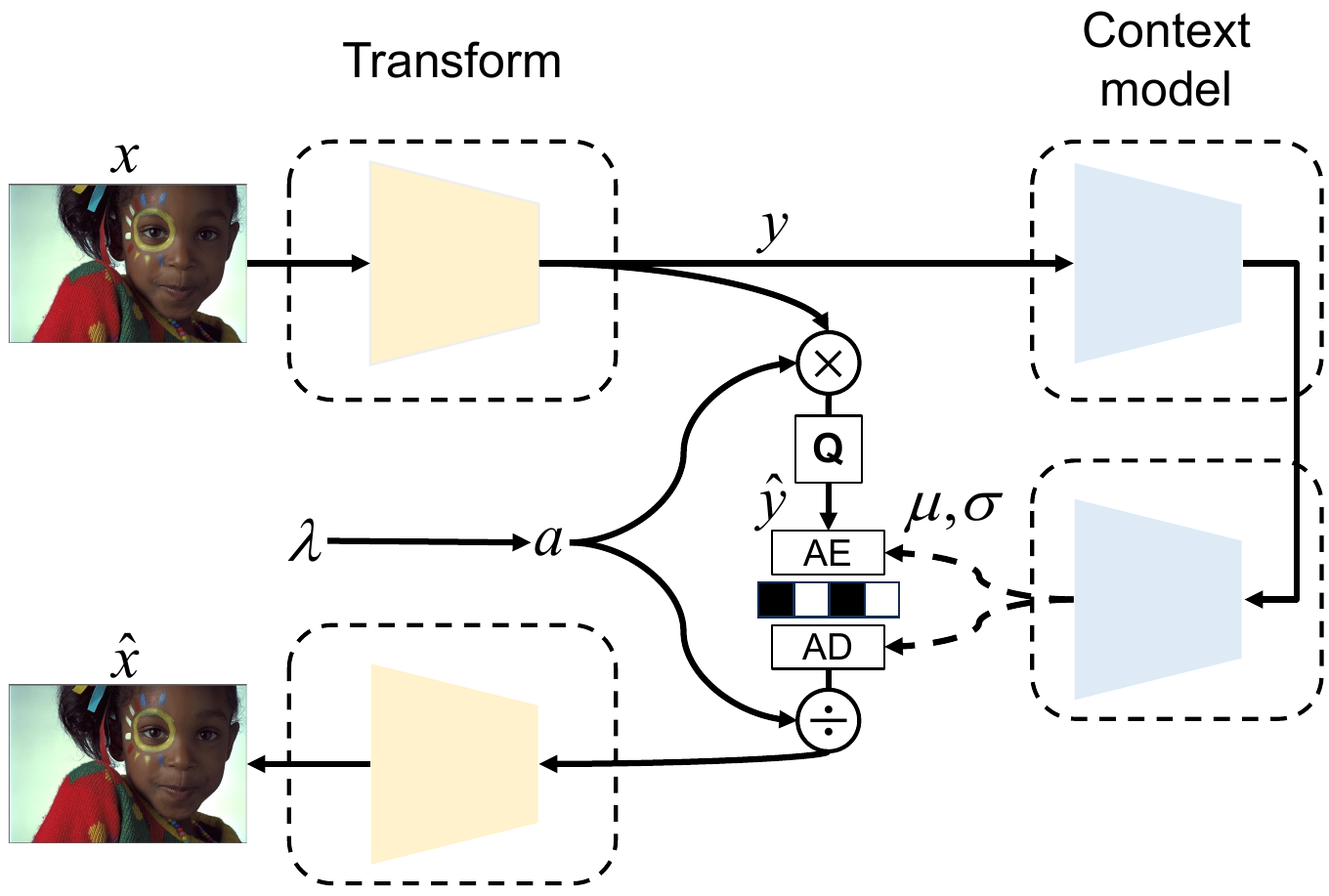}
    \caption{{The framework of baseline model.}} 
    \label{fig:QVRF} 
\end{figure}

We initiate our study by establishing a baseline model utilizing the M\&S Hyperprior~\cite{minnen2018joint} due to its simplicity and efficiency. We adapt it into a variable rate version following QVRF~\cite{Tong2023qvrf}, which involves scaling the latent representation $y$ before encoding and then rescaling it back after decoding. The scaling factor $a$ is initialized based on $\lambda$ using the formula $\sqrt{\frac{\lambda}{\lambda_\text{ref}}}$, with $\lambda_\text{ref}$ set to 0.0018, and it is optimized during training. In the testing phase, we are able to achieve different rates by manually adjusting the value of $a$. The framework of our M\&S Hyperprior-V (M\&S HP-V) is depicted in Fig.~\ref{fig:QVRF}. Our baseline M\&S HP-V (Model A), achieves 14.49\% BD-Rate on the Kodak dataset.

The primary goal of this study is to comprehensively investigate both transforms and context models for LIC. We begin by examining transform modules. Previous studies have demonstrated that networks designed for different purposes can yield satisfactory results in compression~\cite{zhang2024way,duan2023qarv,zhang2024reconfigurable,zhang2023content}. Therefore, our exploration extends beyond compression-related methods to incorporate a diverse array of low-complexity networks from both high-level and low-level vision tasks, including architectures based on convolutions and transformers. 
These networks include Cheng2020~\cite{cheng2020learned}, a LIC method employing the attention mechanism; ELIC~\cite{he2022elic}, a LIC method that performs an excellent balance between speed and compression ability; ConvNeXt~\cite{liu2022convnet}, a modern convolutions network integrating transformer elements for enhanced performance; NAFNet~\cite{chen2022simple}, a method for image restoration that foregoes nonlinear activation, yet delivers high performance; DCNv4~\cite{xiong2024efficient}, a deformable convolution-based large scale foundation model designed for vision tasks; Swin TTC~\cite{zhu2022transformer}, an efficient LIC approach utilizing the Swin transformer; EfficientFormerv2~\cite{li2023rethinking}, a transformer-based vision network that rivals MobileNetV2 in speed while surpassing convolution models in performance; DITN~\cite{liu2023unfolding}, a deployment-friendly transformer for image super-resolution; and TCM~\cite{liu2023learned}, a best-performing LIC method which combines convolution and transformer architectures. 

Next, we replace the transform module of the M\&S HP-V (Model A) with various network architectures. To ensure a fair comparison, we increase the parameters of all models to a similar level. The ``Transform Experiment" part of Fig.~\ref{fig:rdc} and Tab.~\ref{tab:bdc} presents the results of our transform module substitution experiments. Our experiments indicate that certain methods, such as Cheng2020 HP~\cite{cheng2020learned} and EfficientFormerv2 HP~\cite{li2023rethinking}, do not significantly outperform M\&S HP-V (Model A). Despite Cheng2020-GMM rivaling VVC~\cite{cheng2020learned}, and EfficientFormerv2 showing high accuracy with low inference speeds on ImageNet-1K~\cite{li2023rethinking} -- they fall short in our compression tasks across a wide bpp range. In contrast, the best-performing compression method TCM retains robust performance even under Hyperprior settings. Notably, NAFNet HP (Model B) and DCNv4 HP (Model C) emerge as the top performers, thanks to their inherent unique architectural features. Specifically, NAFNet's proficiency in image restoration~\cite{chen2022simple} contributes to its effective handling of compression artifacts, thereby preserving image quality during compression. Similarly, DCNv4's design for large-scale vision tasks equips it with a robust feature extraction capability~\cite{xiong2024efficient}, which is essential for maintaining high-quality image representation.

To gain a deeper understanding of the distinctions between different networks, we adopt the ``\textit{transform coding}~\cite{tc}" view, essentially treating each network as a unique transform and evaluating its effectiveness by measuring the correlation in the latent space. This correlation is measured on $\hat{y}$ as shown in Fig.~\ref{fig:QVRF}. To estimate the de-correlation ability of different transforms, we utilize the latent correlation $\rho_{k \times k}$ presented in~\cite{ali2023towards} and calculate it as follows:
$$\rho_{k \times k}[i] = \mathbb{E}_{x\sim p(x)} \left[ \left( \frac{y_i - \mu_i}{\sigma_i} \right) \left( \frac{y_c - \mu_c}{\sigma_c} \right) \right]$$
where $0 \leq i < k^2$, $k \times k$ is the window size, which is set to 5 in our study and we use the Kodak dataset for calculation. $\mu$ and $\sigma$ refer to the estimated distribution parameters from the context model. $i$ represents the $i$-th element, and $c$ refers to the central point of the window. The window slides across the entire latent space with a stride of 1. After collecting all the windows, we obtain the average of all the windows as the final $\rho_{k \times k}$.

Tab.~\ref{tab:bdc} shows that higher efficiency is associated with lower average correlation. For example, NAFNet HP (Model B) (0.0699 Avg $\rho$, 1.98\% BD-Rate) and DCNv4 HP (Model C) (0.0827 Avg $\rho$, 0.56\% BD-Rate) outperform M\&S HP-V (Model A) (0.1295 Avg $\rho$, 14.49\% BD-Rate) and EfficientFormerv2 HP (0.1448 Avg $\rho$, 14.22\% BD-Rate). Moreover, we find that the effective receptive field (ERF) significantly impacts correlation, with Transformer models featuring larger ERF~\cite{zhu2022transformer} exhibiting higher average correlation. This can be attributed to the calculation of the correlation using a $k \times k$ window. Since Transformers de-correlate over a larger area compared to locally focused convolution models, they exhibit higher local correlation. For instance, Swin HP and ELIC HP demonstrate comparable BD-Rate (7.25\% versus 7.06\%) but differ significantly in their average correlation (0.1127 versus 0.0996). Similarly, while DCNv4 HP (Model C) outperforms NAFNet HP (Model B) in terms of BD-Rate (0.56\% compared to 1.85\%), it exhibits a higher average correlation (0.0827 compared to 0.0699). This discrepancy can be attributed to DCNv4 HP (Model C)'s long-range dependency, which impacts its correlation measurement within the $k \times k$ windows.

Overall, our results on transform modules underscore the importance of strong de-correlation capabilities in achieving efficient compression. We also identify two models, NAFNet HP (Model B) and DCNv4 HP (Model C), be to the better models that achieve a balanced RDC trade-off, as depicted in Fig.~\ref{fig:rdc}. These models form the basis for further context model study.

\begin{table}[htbp]
\centering
\caption{BD-Rate and Measurement.}
\resizebox{0.5\textwidth}{!}{
\begin{tabular}{c|c|c|c|c}
\hline
& \textbf{Model} & \textbf{BD-Rate} & \textbf{Avg $\rho$} &\textbf{Avg Bits}\\ \hline
\multirow{10}{*}{\rotatebox[origin=c]{90}{\parbox{2cm}{\centering Transform \\ Experiments}}} 
& M\&S HP-V (Model A) & 14.49\% & 0.1295 & \multirow{10}{*}{-} \\
& EfficientFormerv2 HP & 14.22\% & 0.1448 & \\
& Cheng2020 HP & 12.99\% & 0.1244 & \\
& Swin HP & 7.25\% & 0.1127 & \\
& ELIC HP & 7.06\% & 0.0996 & \\
& DITN HP & 5.18\% & 0.1095 & \\
& TCM HP & 3.76\% & 0.0997 & \\ 
& ConvNeXt HP & 2.38\% & 0.0839 & \\
& NAFNet HP (Model B) & 1.98\% & 0.0699 & \\
& DCNv4 HP (Model C) & 0.56\% & 0.0827 & \\\hline
\multirow{9}[-12]{*}{\rotatebox[origin=c]{90}{\parbox{2.5cm}{\centering Context Model \\ Experiments}}} 
& NAFNet QARV & -1.45\% &\multirow{3}{*}{-} &0.1936 \\
& NAFNet CHARM & -2.04\% &  &0.1732 \\
& NAFNet SCCTX (Model D) & -3.25\% & & 0.1628 \\ \cline{2-5}
& DCNv4 QARV & -1.98\% &\multirow{3}{*}{-}  & 0.0656\\
& DCNv4 CHARM & -3.33\% &  & 0.0553\\
& DCNv4 SCCTX (Model E) & -4.10\%& &  0.0471\\ \hline
\end{tabular}} 
\begin{tablenotes}
  \item BD-Rate is calculated on the Kodak dataset.
  \item The average $\rho$ is obtained by taking the mean of $\rho_{k \times k}$. 
  \item The average bits for the NAFNet series are calculated using kodim14, while the average bits for the DCNv4 are calculated using kodim23. $\lambda=0.0018$
  \end{tablenotes}
  \vspace{-5mm}
  \label{tab:bdc}
\end{table}

%% file: context_study.tex
\section{Context Model Study}
\label{sec:Context}
Regarding context models, we pick three representative low-complexity context models. These models are: (1) the Channel-wise Autoregressive Models (CHARM)~\cite{minnen2020channel}, which models channel relationships; (2) the Space-Channel Context Models (SCCTX)~\cite{he2022elic}, which considers both channel and spatial dependencies; and (3) the Quantization-Aware ResNet VAE (QARV)~\cite{duan2023qarv}, which generalizes previous autoregressive methods. Notably, because QARV inherently supports variable rate compression, we do not apply additional modifications to facilitate variable rate capabilities.

We implement these context models based on NAFNet HP (Model B) and DCNv4 HP (Model C). We keep the original design for CHARM, whereas, for SCCTX, we increase its complexity by expanding the number of channels to achieve similar parameters. For QARV, we opt for a four latent variables version to regulate the parameters. The results on the context models, illustrated in Fig.~\ref{fig:rdc} and Tab.~\ref{tab:bdc}, show that all the explored models outperform the VVC Intra coding mode. A consistent trend is also observed, with SCCTX surpassing CHARM, and CHARM outperforming QARV. For instance, DCNv4 SCCTX (Model E) gains -4.10\% BD-Rate $\textgreater$ DCNv4 CHARM achieves -3.33\% BD-Rate $\textgreater$ DCNv4 QARV yields -1.98\% BD-Rate. 

This trend aligns with our expectations, considering the innovative integration of a checkboard model and a channel-wise autoregressive model in SCCTX. SCCTX also considers the issue of uneven channel information distribution in CHARM. Consequently, SCCTX, as a refined iteration of CHARM, naturally showcases enhanced performance. On the other hand, QARV's performance lags behind, partly because of its lack of a predefined coding pattern, which complicates convergence under the same training conditions.

We further illustrate how different bits allocation is applied under the same transform network (even though the weights are different, it can be stated that the same network structure possesses an equivalent capacity~\cite{li2018visualizing}.). Fig.~\ref{fig: bits} shows that smoother image areas require fewer bits (darker color), while complex textured regions demand more bits (brighter color). Additionally, the utilization of advanced context models enables more accurate estimations of joint model distribution parameters ($\mu$ and $\sigma$)~\cite{yang2023introduction}. This is facilitated by the predefined coding pattern in the context model, which aligns with the information distribution~\cite{he2022elic}, resulting in an overall reduction in bit requirements (refer to Tab.\ref{tab:bdc} and Fig~\ref{fig: bits}).

Among these exploratory models, two are distinguished for their exemplary RDC tradeoff, NAFNet SCCTX (Model D) and DCNv4 SCCTX (Model E). Model E, designated as our final model, demonstrates exceptional efficiency, achieving -4.10\% BD-Rate through the synergy of the DCNv4 transform and SCCTX context model. The low complexity and superior performance of our models highlight the potential of integrating advanced context models with best-performing transform networks for efficient LIC models.

\newcommand{\mynewspace}{0.23}
\begin{figure*}[htbp]
  \centering 
  \begin{minipage}[b]{1\linewidth} 
  \centering
   \subfigure[kodim14]{
    \begin{minipage}[b]{0.1962\linewidth} 
      \centering
      \includegraphics[width=\linewidth]{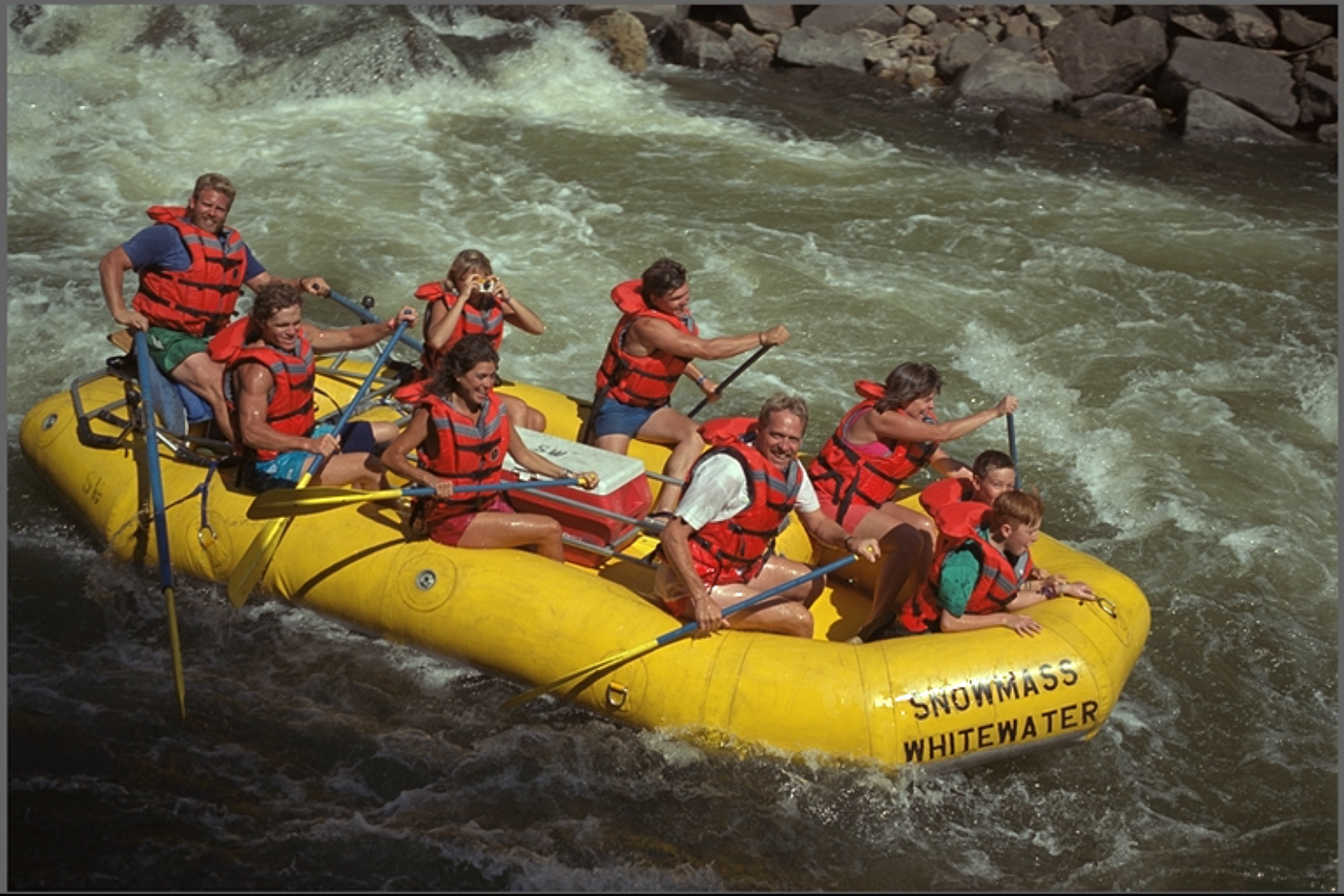}\\
    \end{minipage}
  }
      \subfigure[NAFNet HP]{
    \begin{minipage}[b]{\mynewspace\linewidth} 
    \label{subfig:NAFHP}
      \centering
      \includegraphics[width=\linewidth]{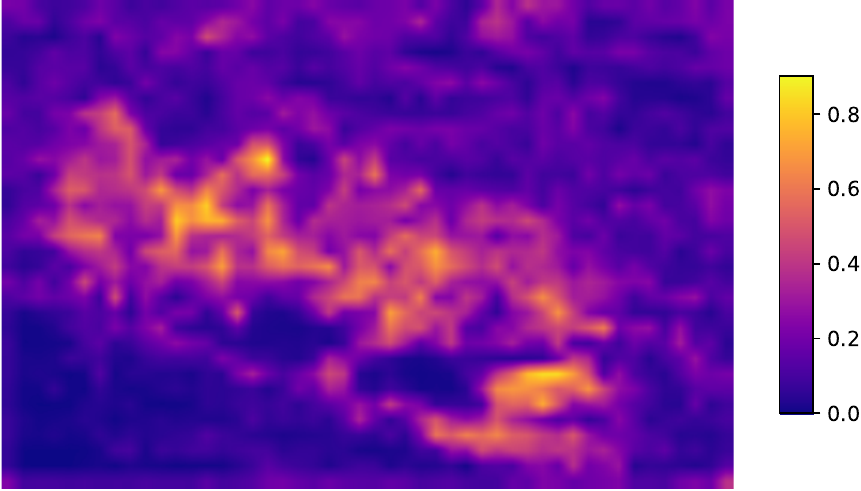}\\
    \end{minipage}
  }
   \subfigure[NAFNet CHARM]{
    \begin{minipage}[b]{\mynewspace\linewidth} 
      \centering
      \includegraphics[width=\linewidth]{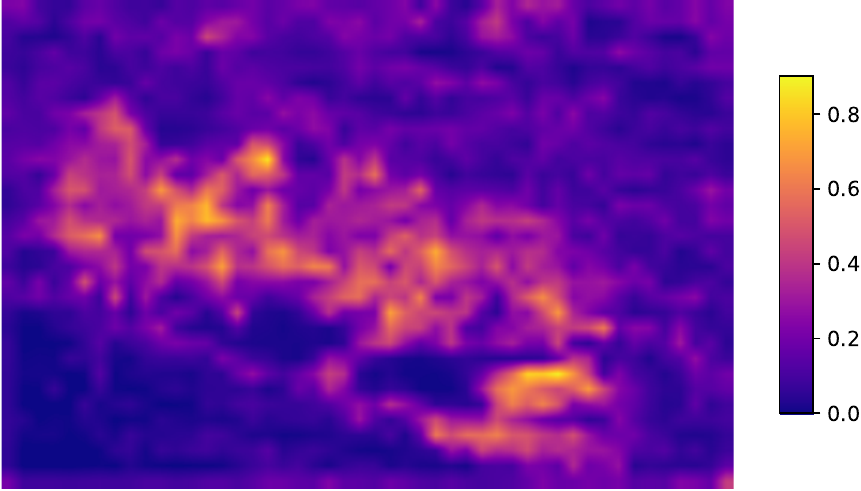}\\
    \end{minipage}
  }
   \subfigure[NAFNet SCCTX]{
    \begin{minipage}[b]{\mynewspace\linewidth} 
    \label{subfig:NAFSCCTX}
      \centering
      \includegraphics[width=\linewidth]{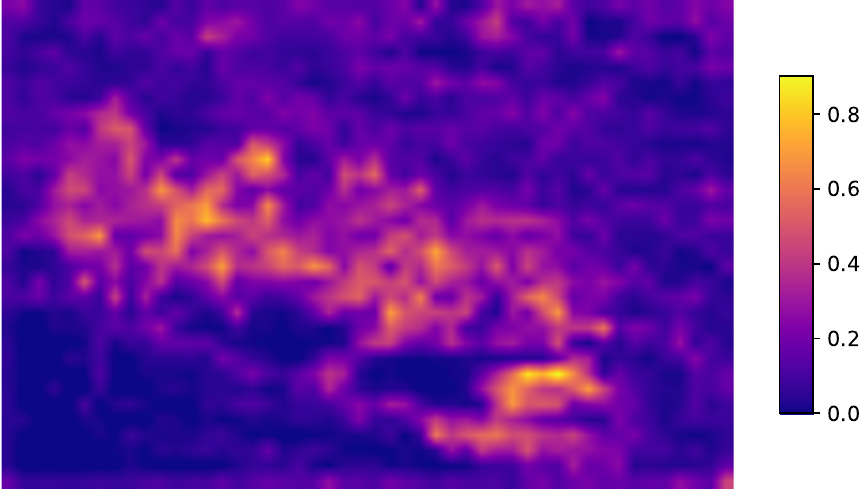}\\
    \end{minipage}
  }
      \subfigure[kodim23]{
    \begin{minipage}[b]{0.1962\linewidth} 
      \centering
      \includegraphics[width=\linewidth]{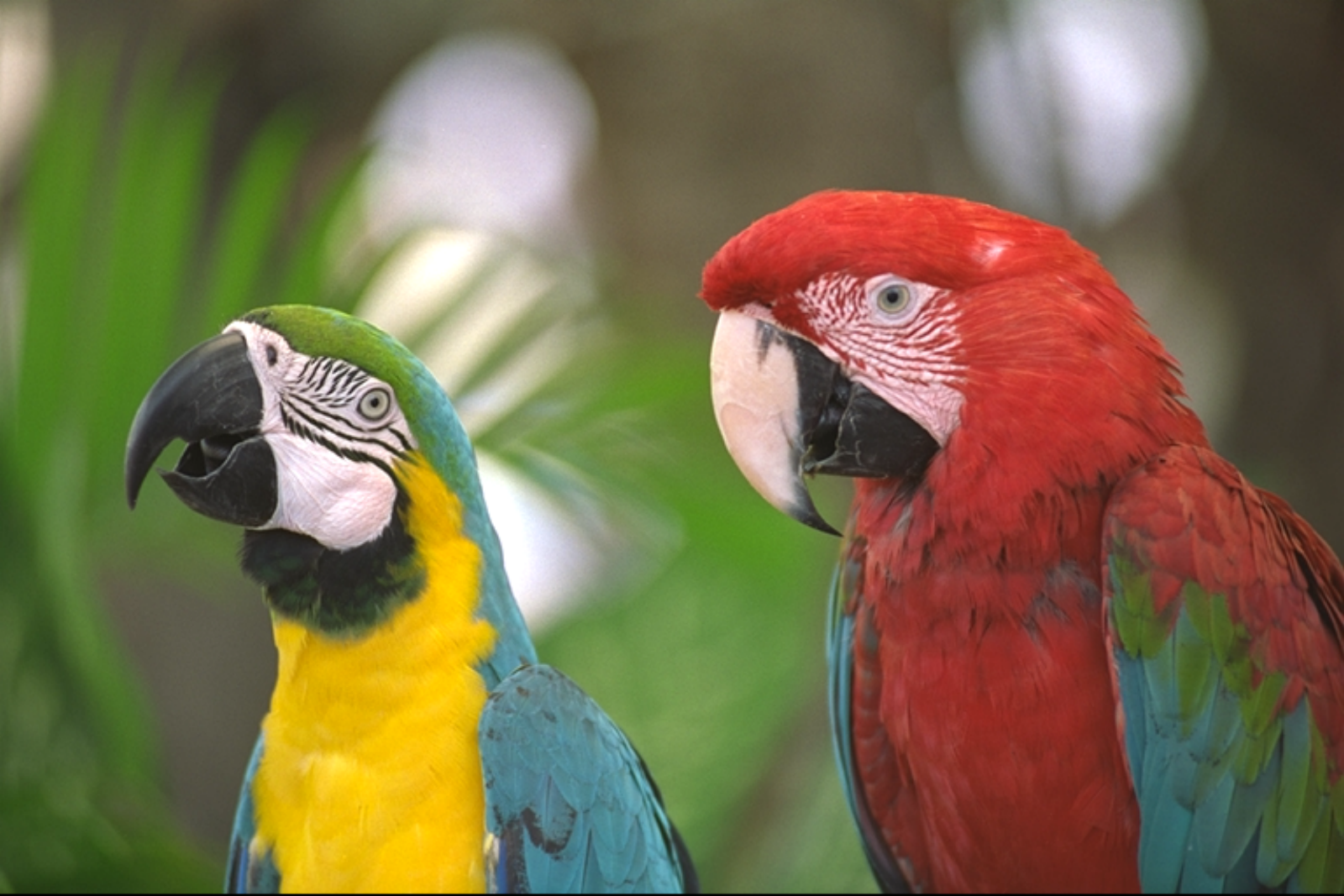}\\
    \end{minipage}
  }
  \subfigure[DCNv4 HP]{
    \begin{minipage}[b]{\mynewspace\linewidth} 
    \label{subfig:DCNHP}
      \centering
      \includegraphics[width=\linewidth]{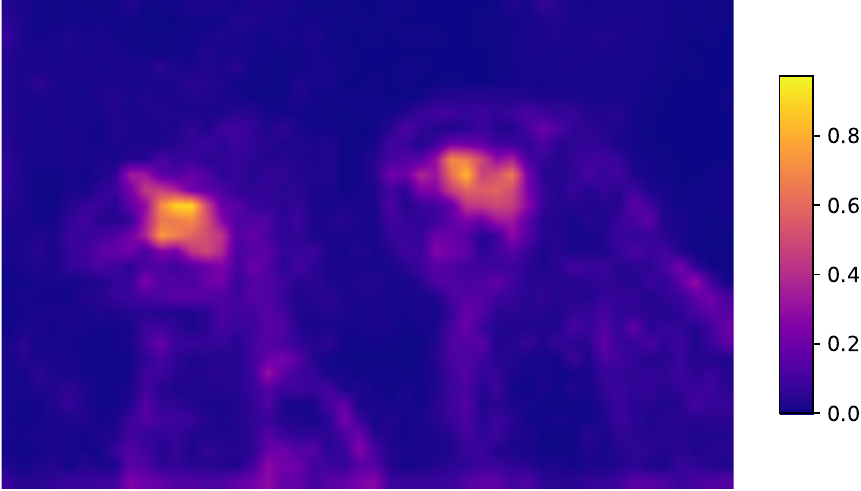}\\
    \end{minipage}
  }
   \subfigure[DCNv4 CHARM]{
    \begin{minipage}[b]{\mynewspace\linewidth} 
      \centering
      \includegraphics[width=\linewidth]{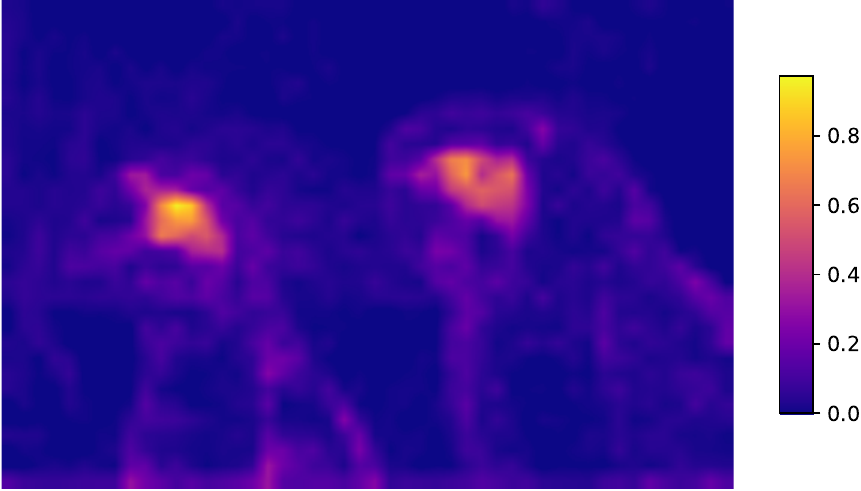}\\
    \end{minipage}
  }
   \subfigure[DCNv4 SCCTX]{
    \begin{minipage}[b]{\mynewspace\linewidth} 
    \label{subfig:DCNSCCTX}
      \centering
      \includegraphics[width=\linewidth]{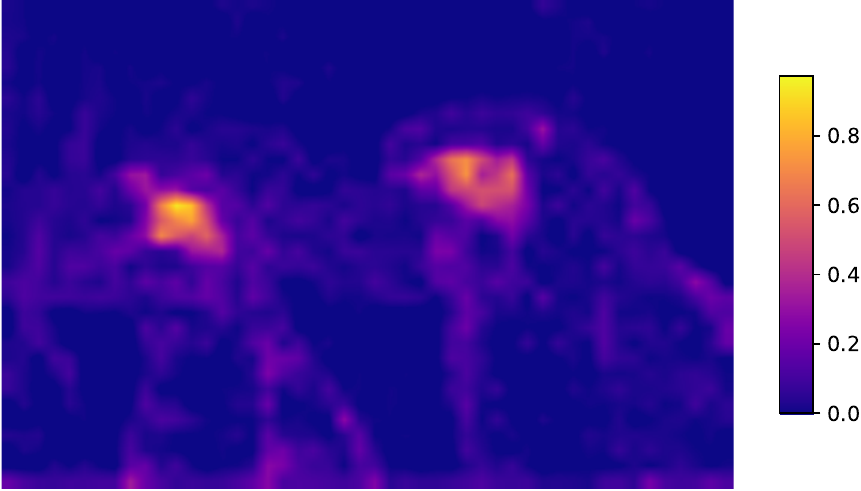}\\
    \end{minipage}
  }

   \end{minipage}
\vspace{-4mm}
  \caption{{Visualization of the bits allocation using different context models. $\lambda = 0.0018$.} Note: QARV is fundamentally different from other context models, so this visualization of bits allocation is not applicable to QARV.\vspace{-2mm}}
  \label{fig: bits}
\end{figure*}

%% file: experimental_setup.tex
\vspace{-2mm}
\section{Experimental setup}
\label{sec:expset}
\vspace{-2mm}

\subsection{Training} We use COCO 2017 Trainset~\cite{lin2014microsoft} as our training sets, which comprises 118,287 images with a resolution of $640\times420$ pixels. We randomly cropped $256\times256$ patches for training. To cover a wide range of bit-rates, we define our Lagrange multiplier set $\Lambda$ as \{0.0018, 0.0035, 0.0067, 0.0130, 0.0250, 0.0483, 0.0932, 0.1800, 0.3600, 0.7200, 1.4400\}. Following QVAF~\cite{Tong2023qvrf}, the training process consists of three phases. In the first phase, we use a $\lambda$ value of 1.4400 and apply uniform noise as quantization for 100 epochs. During the second phase, $\lambda$ is randomly selected from $\Lambda$, and uniform noise quantization is applied for 80 epochs. In the final phase, we again randomly select $\lambda$ from $\Lambda$ but using a straight-through estimator (STE)~\cite{bengio2013estimating} approximation for quantization for 70 epochs. Each model is trained using the Adam optimizer, with a batch size of 32 and an initial learning rate of 2e-4. We employed the ReduceLRonPlateau learning rate scheduler, configured with a patience of 10 epochs and a reduction factor of 0.5. It is important to note that each transition between training phases involves resetting the learning rate to 2e-4. All models are implemented and trained using Pytorch.

\subsection{Testing} Three widely-used benchmark datasets, including Kodak\footnote{\url{https://r0k.us/graphics/kodak/}}, Tecnick\footnote{\url{https://tecnick.com/?aiocp\%20dp=testimages}}, and CLIC 2022\footnote{\url{http://compression.cc/}}, are used to evaluate the performance of the proposed models. Also, following previous work~\cite{minnen2018joint,minnen2020channel,he2022elic} and community discussions\footnote{\url{https://groups.google.com/g/tensorflow-compression/c/LQtTAo6l26U/m/mxP-VWPdAgAJ}}, we do not encode $\lceil y \rfloor$ for entropy coding, instead, we encode each $\lceil y-\mu \rfloor$. When reverting the encoded symbols, we revert them to $\lceil y-\mu \rfloor + \mu$, which enables the single Gaussian entropy model to yield better results.

%% file: complexity.tex
\vspace{-0.3cm}
\section{Complexity analysis}
\label{sec:com}

\begin{table*}[h]
    \centering
    \small
    \caption{Computational Complexity and BD-Rate}
    \begin{tabular}{c|c|c|c|c|c|c|c}\hline\hline
        \multirow{2}{*}{Method} & \multirow{2}{*}{MACs/pixel} & \multirow{2}{*}{Params.} & \multicolumn{2}{c|}{Latency (GPU)} & \multicolumn{3}{c}{BD-Rate (\%) w.r.t. VTM 18.0} \\
        \cline{4-8}  &  &  & Enc. & Dec.  & Kodak & Tecnick & CLIC2022  \\\hline
        \multicolumn{8}{c}{Hyperprior-based Method}\\\hline
        M\&S Hyperprior~\cite{minnen2018joint}&438K &98.4M  & 0.065s & 0.069s & 17.05\% & 26.08\% & 29.12\% \\
        M\&S HP-V (Model A) &438K &17.56M & 0.064s & 0.067s & 14.49\% & 18.33\% & 21.48\% \\
        NAFNet HP (Model B) &799K &23.57M & 0.080s & 0.066s & 1.98\% & 0.58\% & 0.90\% \\
        DCNv4 HP (Model C) &1142K &27.90M & 0.096s & 0.081s & 0.56\% & -0.96\% & -0.45\% \\\hline
        \multicolumn{8}{c}{Spatial Autoregressive-based Method}\\\hline
        Cheng2020~\cite{cheng2020learned}&926K &115.3M & 3.121s  & 6.823s & 3.94\% & 5.93\% & 8.30\% \\\hline
        \multicolumn{8}{c}{Advanced Context Model-based Method}\\\hline
        STF~\cite{zou2022devil}& 509K  &599.1M & 0.160s & 0.147s & -2.52\% & -2.35\% & -1.18\% \\
        TCM-S~\cite{liu2023learned}& 539K &271.1M & 0.271s & 0.162s & -5.52\% & -8.15\% & -8.53\% \\ 
        NAFNet SCCTX (Model D) &869k &40.62M & 0.247s & 0.134s & -3.25\% & -4.44\% & -4.22\% \\
        DCNv4 SCCTX (Model E) &1212K &44.95M & 0.264s & 0.148s & -4.10\% & -5.91\% & -5.60\% \\
        \hline\hline
    \end{tabular}
\begin{tablenotes}
  \item {\bf Test Conditions}: Nvidia A40 GPU. The enc./dec. time is averaged over all 24 images in Kodak, including entropy enc./dec. time.
  \end{tablenotes}
  \label{tab:bdrate}%
\end{table*}%

\begin{figure*}[h] 
\newcommand{\mywidth}{0.3}
\centering 
\subfigure[Kodak]{\includegraphics[width=\mywidth\linewidth]{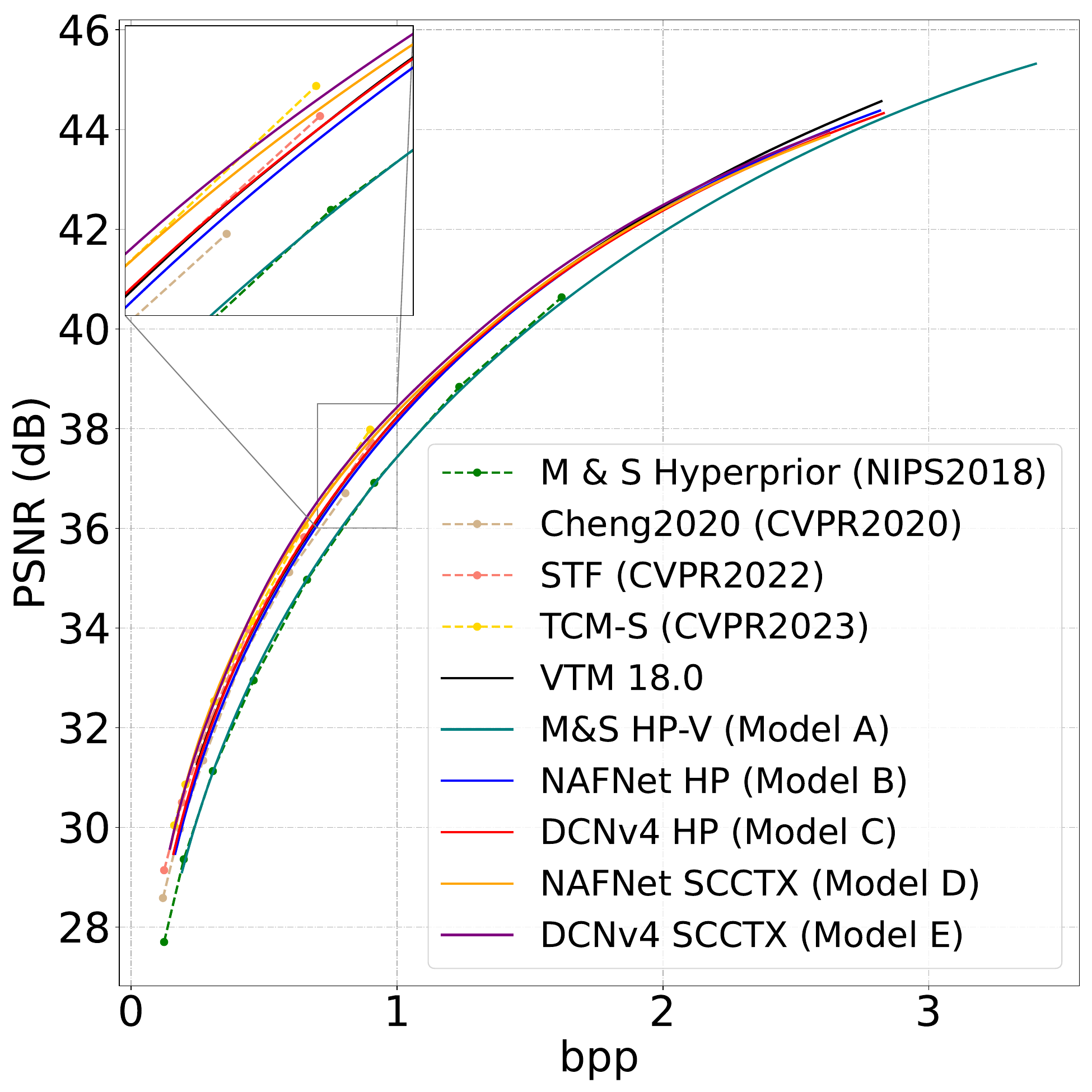}\label{subfig:kodak}} 
\subfigure[Tecnick 1200x1200]{\includegraphics[width=\mywidth\linewidth]{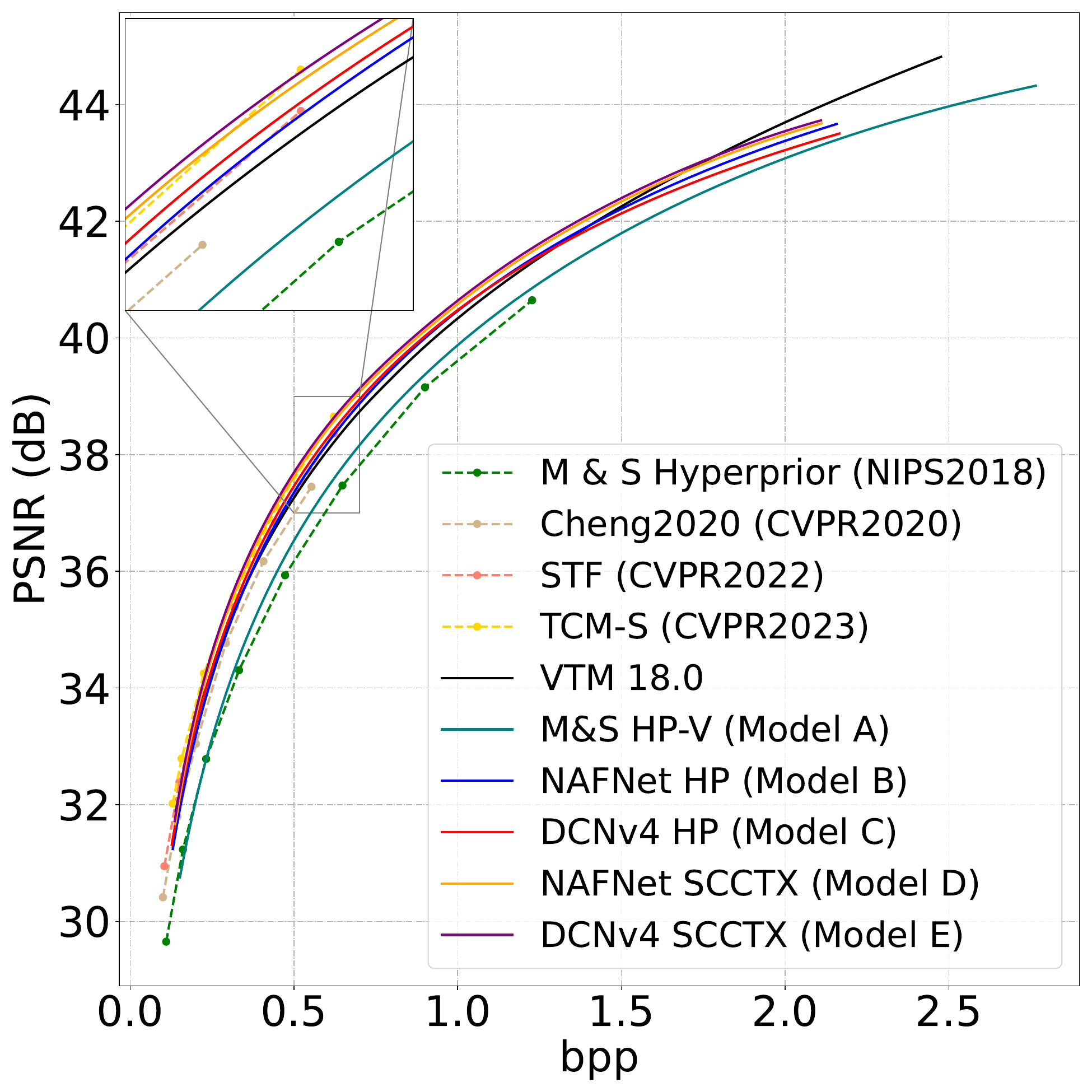}\label{subfig:Tecnick}}
\subfigure[CLIC 2022]{\includegraphics[width=\mywidth\linewidth]{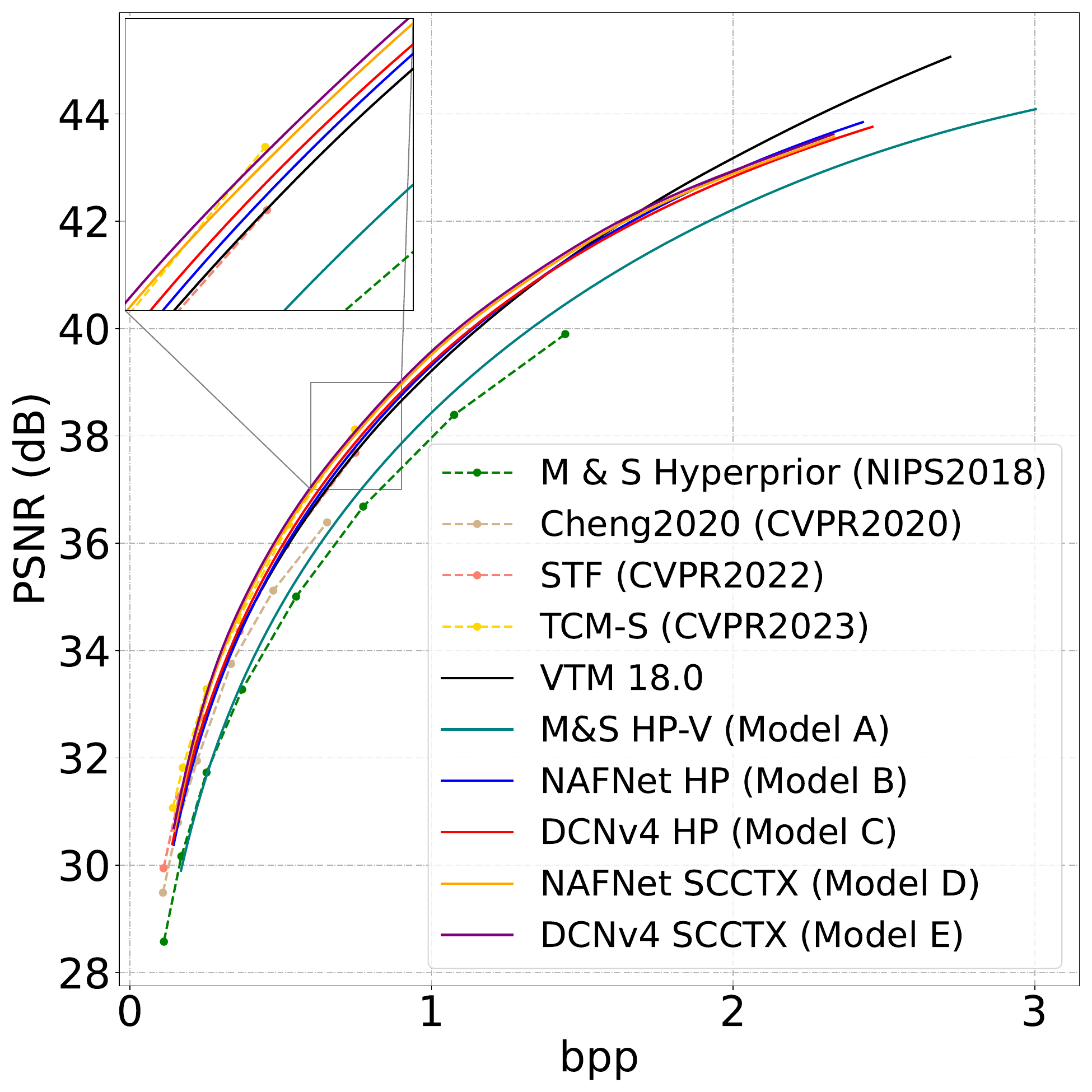}\label{subfig:CLIC}}
\vspace{-2mm}
\caption{\textbf{RD curves of various methods. }{\it Please zoom in for more details}.\vspace{-4mm}} 
\label{fig:mse_result} 
\end{figure*}

The objective of our study is to develop a series of models that offer the best RDC trade-off. Our analysis (Section \ref{sec:Transform} and Section \ref{sec:Context}) has shown the best models are M\&S HP-V (Model A), NAFNet HP (Model B), DCNv4 HP (Model C), NAFNet SCCTX (Model D), and DCNv4 SCCTX (Model E). In this section, we evaluate the complexity of these models by comparing them to prominent LIC methods such as M\&S Hyperprior~\cite{minnen2018joint}, Cheng2020~\cite{cheng2020learned}, STF~\cite{zou2022devil}, and TCM-S~\cite{liu2023learned}. 

To evaluate the performance, we utilize the Bj{\o}ntegaard delta bit-rate (BD-Rate)~\cite{BDrate} to assess bit-rate reduction while maintaining image quality as measured by PSNR. We use VTM-18.0\footnote{\url{https://vcgit.hhi.fraunhofer.de/jvet/VVCSoftware_VTM}} All Intra as the anchor for calculating BD-Rate. The complexity is measured using total parameters (The total parameters of the fixed rate method are obtained by summing up model parameters for all bpp.), multiply-accumulate operations per pixel (MACs/pixel), encoding time (Enc.), and decoding time (Dec.) on GPU.

Our models demonstrate impressive performance across various datasets, considering their complexity as shown in Fig.~\ref{fig:mse_result} and Tab.~\ref{tab:bdrate}. M\&S HP-V (Model A), an improved version of M\&S Hyperprior with QRAF and updated training scripts achieves 2.56\%, 7.75\%, and 7.64\% higher BD-Rate gains than  M\&S Hyperprior for the Kodak, Tecnick, and CLIC2022 datasets, respectively, with only 17\% of the total parameters of the original M\&S Hyperprior. NAFNet HP (Model B) and DCNv4 HP (Model C), both based on the Hyperprior framework, outperform Cheng2020, which employs a Spatial Autoregressive model. For example, Cheng2020 achieves 3.94\% BD-Rate on the Kodak dataset, while DCNv4 HP (Model C) achieves 0.56\% BD-Rate. In addition, DCNv4 HP (Model C) has significantly fewer total parameters (27.90M) compared to Cheng2020 (115.3M), highlighting its efficiency. NAFNet SCCTX (Model D) and DCNv4 SCCTX (Model E), which incorporate advanced context models, achieve results on par with best-performing methods STF and TCM-S. DCNv4 SCCTX (Model E), for instance, achieves -4.10\% BD-Rate, -5.91\% BD-Rate, and -5.60\% BD-Rate on Kodak, Tecnick, and CLIC2022, respectively, outperforming STF despite STF having significantly larger parameters of 599.1M.

In terms of MACs/pixel, our Models A, B, C, D, and E all have higher MACs/pixel. However, this metric solely reflects the number of multiplications and additions and does not directly correlate with increased latency, as supported by prior research~\cite{minnen2023advancing}. For example, DCNv4 HP (Model C) has 1142K MACs, compared to TCM-S's 539K, yet DCNv4 HP (Model C)'s decoding time is only 0.081 seconds, half that of TCM-S. Similarly, DCNv4 SCCTX (Model E), with 1212K MACs, has a decoding time competitive with STF and slightly faster than TCM-S.

Regarding encoding and decoding latency, all our models are competitive with existing methods, which is largely influenced by the choice of context model. Models A, B, and C, which are based on the Hyperprior framework, have similar encoding and decoding times to M\&S Hyperprior. On the other hand, Models D and E, which use more complex context models, have similar times to STF and TCM-S, showcasing the efficiency of our models across various computational and performance metrics.

%% file: conclusion.tex
\vspace{-5mm}
\section{Conclusion}
\vspace{-4mm}
\label{sec:con}
In this study, we investigate various architectures for image compression, with an emphasis on achieving the best RDC trade-off. We find that transform modules with strong decorrelation capabilities and context models with accurate modeling abilities are the cornerstones of strong LICs. Our results demonstrate that by carefully selecting and integrating transform and context models, it is possible to achieve strong compression performance while reducing complexity. The proposed final model, which combines DCNv4 with the SCCTX context model, demonstrates exceptional efficiency, highlighting the potential for designing highly efficient models with significantly reduced complexity.

\textbf{Limitations and future work.} It is important to note that this study primarily relies on empirical investigation and lacks a theoretical analysis of the rate-distortion-complexity trade-off. Furthermore, we have not explored other methods for reducing complexity, such as knowledge distillation, model quantization, and pruning, which will be done in future work.